\documentclass[12pt,a4paper]{article}
\usepackage{amsmath,amscd}
\usepackage{amssymb}
\usepackage{bm}
\usepackage{cite}  %%% converts [1,2,3,4,5,6] to [1--6] in citations
\usepackage{comment}
\usepackage{graphicx}
\usepackage{hyperref}
\usepackage{makeidx}
\usepackage{mathpazo} % typeface
\usepackage{mathtools}
\usepackage{multimedia}
\usepackage{slashed}
\usepackage[small,nohug,heads=vee]{diagrams}
\usepackage[usenames,dvipsnames]{xcolor}
\usepackage{xfrac}

\def\dint{\mathop{\displaystyle \int}}

\definecolor{light-gray}{gray}{0.80}
\definecolor{dRed}{RGB}{0, 0, 0}
\diagramstyle[labelstyle=\scriptstyle]

\newcommand\eq{=&&\hspace{-18pt}}

\newcommand\strt[1]{\rule[-#1pt]{0pt}{#1pt}}

\usepackage{fancyhdr}
 \renewcommand{\maketitle}{
     \begin{center}
       \Large
         {\bf The Particle as a Statistical Ensemble of Events in~Stueckelberg--Horwitz--Piron Electrodynamics}
         \vskip .3 true cm
       \small
         Martin Land \\
         \vskip .3 true cm
         Department of Computer Science \\
         Hadassah College \\
         37 HaNevi'im Street, Jerusalem \\
 email: martin@hac.ac.il
       \end{center}
       \vskip .5 true cm
 }
\input{tcilatex}

\begin{document}
\title{}
\author{}
\maketitle

% \title{Title}
% \author{Martin Land}
% \address{Department of Computer Science, Hadassah College, 37 HaNeviim Street, Jerusalem}
% \ead{martin@hac.ac.il}

%
\begin{abstract}
In classical Maxwell electrodynamics, charged particles following
deterministic trajectories are described by currents that induce fields,
mediating interactions with other particles.  Statistical methods are used when
needed to treat complex particle and/or field configurations.  
In~Stueckelberg--Horwitz--Piron (SHP) electrodynamics, the classical trajectories are
traced out dynamically, through the evolution of a 4D spacetime event $x^\mu
(\tau)$ as $\tau$ grows monotonically.  Stueckelberg proposed to formalize the
distinction between coordinate time $x^0 = ct$ (measured by laboratory clocks)
and chronology $\tau$ (the temporal ordering of event occurrence) in order to
describe antiparticles and resolve problems of irreversibility such as
grandfather paradoxes.  Consequently, in SHP theory, the elementary object is not
a particle (a 4D curve in spacetime) but rather an event (a single point along
the dynamically evolving curve).  Following standard deterministic methods in
classical relativistic field theory, one is led to Maxwell-like field equations
that are \hbox{$\tau$-dependent} and sourced by a current that represents a
statistical ensemble of instantaneous events distributed along the trajectory.
The width $\lambda$ of this distribution defines a correlation time for the
interactions and a mass spectrum for the photons emitted by particles.  As
$\lambda$  becomes very large, the photon mass goes to zero and the field
equations become $\tau$-independent Maxwell's equations.  Maxwell theory thus
emerges as an equilibrium limit of SHP, in which $\lambda$ is larger than any
other relevant time scale.  Thus, statistical mechanics is a fundamental
ingredient in SHP electrodynamics, and its insights are~required to give meaning
to the concept of a particle.
\end{abstract}

\baselineskip7mm 
\parindent=0cm \parskip=10pt

\section{Introduction}
In developing his interpretation of antiparticles as particles travelling
backward in time, Stueckelberg \cite{Stueckelberg-1,Stueckelberg-2} hoped to demonstrate that
pair creation/annihilation processes would appear naturally in a thoroughly
deterministic and classical relativistic Hamiltonian mechanics, once such a~formalism
could be properly constructed.  In this picture, a particle is the worldline traced
out by a~spacetime event $x^\mu\left(\tau\right)$, for $\mu = 0,1,2,3$,
that evolves dynamically as
the Poincar\'{e} invariant parameter proceeds monotonically from $\tau =-\infty
$ to $\tau =\infty $.  Stueckelberg argued that pair annihilation is observed
when the coordinate time $x^0(\tau)$ reverses direction, because for some values
$x^0_2 > x^0_1$ on the~laboratory clock there will be two solutions to
$x^0(\tau)=x^0_1$ but no solution to $x^0(\tau)=x^0_2$.  He~proposed a covariant
evolution equation of the form
\begin{equation}
\frac{D }{D\tau} \dot x^\mu = 
\dfrac{d^{2}x^{\mu }}{d\tau ^{2}}+\Gamma _{\nu \rho }^{\mu }\dfrac{dx^{\nu }
}{d\tau }\dfrac{dx^{\rho }}{d\tau }=\frac{e}{M}\left[ F^{\mu \nu }(x) g_{\nu \rho }\dfrac{
dx^{\rho }}{d\tau }+G^{\mu }(x) \right]
\label{10}
\end{equation}
in which $\Gamma _{\nu \rho }^{\mu }$ is the Christoffel connection, $F^{\mu \nu
}$ is the electromagnetic field tensor and
the vector field $G^\mu(x)$ is required to overcome the 
mass-shell constraint
\begin{equation}
\frac{D}{D\tau} \left( \frac12 M \dot x^2 \right) = M \dot x_\mu \frac{D \dot x^\mu}{D\tau} = e \dot x_\mu
G^\mu(x)
\underset{G^\mu \ \rightarrow \ 0 }{\mbox{\
}\xrightarrow{\hspace*{1.5cm}}\mbox{\ }} 0 % .
\end{equation}
of standard relativity.  
By keeping $\dot x^2$ from changing sign, this constraint prevents the event from
crossing the space-like region that separates future-oriented trajectories from
past-oriented trajectories.  However, Stueckelberg was not satisfied that he
could justify a Hamiltonian $K$ that produces his evolution equation in flat
spacetime from the unconstrained symplectic equations   
\begin{equation}
\dfrac{dx^{\mu }}{d\tau }=\dot{x}^{\mu }=\dfrac{\partial K}{\partial p_{\mu } }
\qquad
\dfrac{dp^{\mu }}{d\tau }=\dot{p}^{\mu }=-\dfrac{ \partial K}{\partial x_{\mu }}
\label{80}
\end{equation}
and instead continued his program in quantum mechanics, where, as in Feynman's
spacetime diagrams, the event may tunnel probabilistically across the space-like
region.

%%% footnote moved to new paragraph
{{It should be noted that the~Feynman--Stueckelberg interpretation
of antiparticles relies on the Standard Model, in~which all matter is composed
of charged quarks and leptons subject to the strong and electroweak forces
described by non-Abelian gauge theories.  Thus, while the neutrino is
electrically neutral, it carries the~weak nuclear charge and is distinguished
from the antineutrino by a time reversal produced through a~generalization of}
(\ref{10}) {to the weak nuclear force.  Moreover, neutral mesons are~understood
as quark bound states which are structurally symmetric under time reversal.  Thus,
under the~electromagnetic force, the individual constituents of the $D0$ meson (the bound
state $u\bar{c}$ formed from an up quark and a charmed antiquark) may undergo time
reversal separately, resulting~in~the~$\overline{D0}$ bound state $\bar{u}c$.}}

Horwitz and Piron \cite{HP} returned to some of these questions in constructing a
canonical relativistic mechanics for the two-body problem.  
Introducing an invariant scalar interaction of the type
\begin{equation}
K = \frac{p^2_1}{2M_1} + \frac{p^2_2}{2M_2} + V\left(\vert x_1 - x_2 \vert
\right)
\label{K_HP}
\end{equation}
and replacing 
\begin{equation}
R = \sqrt { \left( {\mathbf x}_1 - {\mathbf x}_2 \right)^2 } \qquad \longrightarrow
\qquad \rho = \sqrt{\left( {\mathbf x}_1 - {\mathbf x}_2 \right)^2 - c^2 \left( t_1
- t_2 \right)^2 }
\end{equation}
in the argument of nonrelativistic scalar potentials, Horwitz et al.\ found
solutions for relativistic generalizations of the standard central force
problems, including quantum mechanical potential scattering and bound states
\cite{bound-1,bound-2,bound-3,bound-4,bound-5}. 
Examination of radiative transitions \cite{radiative-1,radiative-2,radiative-3} associated
with these bound states suggests that the scalar interaction $V$ is required
along with the four-vector potential $A^\mu$ in order to account for known
phenomenology.  Although the electromagnetic tensor interaction produced by
$F^{\mu\nu}$ leaves individual particle masses invariant, the scalar
interaction, which by way of~(\ref{80}) leads to the vector field $G^\mu =
-\partial K / \partial x_\mu$ proposed by Stueckelberg, permits mass exchange 
in~such~transitions. 

A related issue arises in defining the interaction picture in quantum field
%%% definition inserted
theory (QFT), required for the Dyson time-ordered perturbation expansion.  On the one
hand, the Haag theorem proves that any field obtained by unitary transformation
of a free field must itself be a free field.  On~the other hand, the interacting
fields of perturbation theory are defined by acting on free fields with a
unitary transformation generated by an interaction Hamiltonian.  To resolve this
apparent contradiction, Seidewitz has observed \cite{haag} that Haag's proof
relies on transformations generated by a~Hamiltonian which is the 0-component of
a four-vector and parameterized by $x^0$.  He demonstrates that the~construction
of standard QFT in the Stueckelberg framework leads to an interaction picture
obtained by acting on free fields with a unitary transformation generated by a
scalar Stueckelberg Hamiltonian, as for example in Equation (\ref{S-Ham}) below,
and parameterized by $\tau$.  Because this generator is~invariant under Lorentz
transformations, a crucial step in the Haag theorem is inapplicable, and 
the~no-go result is averted.  It is worth noting that even if the Stueckelberg
Hamiltonian is $\tau$-independent, the~interaction picture Hamiltonian will
depend on $\tau$ and so permit mass exchange among particles and fields (just as
a $t$-dependent nonrelativistic Hamiltonian permits nonconservation of energy). 

Sa'ad, Horwitz, and Arshansky \cite{saad} found a more fundamental justification
for the scalar field by studying the gauge invariance associated with
(\ref{K_HP}).  It has been shown \cite{beyond} that the most general classical
interaction consistent with the unconstrained quantum commutation relations 
\begin{equation}
\left[ x^{\mu },x^{\nu }\right] =0\qquad m\left[ x^{\mu },\dot{x}^{\nu }
\right] =-i\hbar g^{\mu \nu }\left( x\right) \qquad \mu,\nu = 0,1,2,3
\label{550}
\end{equation}
is given by Stueckelberg's evolution Equation (\ref{10}) with the substitutions
\begin{equation}
F^{\mu\nu} (x) \rightarrow f^{\mu\nu} (x,\tau) \qquad \qquad
G^\mu (x) \rightarrow f^{5\mu} (x,\tau) ,
\label{fields}
\end{equation}
and that in flat Minkowski space, this system is equivalent to the Lagrangian
\begin{equation}
L = \dot{x}^{\mu }p_{\mu }-K=\dfrac{1}{2}M\dot{x}^{\mu }\dot{x}_{\mu }+\dfrac{e}{c}
\dot{x}^{\mu }a_{\mu }(x,\tau)+\dfrac{e}{c} \phi (x,\tau) 
\label{lagrangian}
\end{equation}
where the electromagnetic field strength tensor
\begin{equation}
f^{\mu\nu} = \partial^\mu a^\nu - \partial^\nu a^\mu \qquad 
f^{5\mu} = \partial_\tau a^\mu - \partial^\mu \phi
\label{f-def}
\end{equation}
is derived from {\em five} potentials.

This classical Lagrangian is unique up to the $\tau$-dependent gauge transformations
\begin{equation}
a_{\mu }(x,\tau ) \rightarrow  a_{\mu }(x,\tau)+\partial_{\mu }\Lambda (x,\tau )
\qquad 
\phi(x,\tau ) \rightarrow \phi(x,\tau )+\partial _{\tau }\Lambda (x,\tau )
\label{gauge_1}
\end{equation}
and the associated quantum mechanics
\begin{equation}
i\hbar \partial _{\tau }\psi \left(x,\tau \right) = K \psi \left(x,\tau \right)
=\left[ \frac{1}{2M}\left(p^{\mu }-\frac{
e}{c}a^{\mu }\right)\left(p_{\mu }-\frac{e}{c}a_{\mu }\right)-\frac{e}{c}\phi 
\right] \psi \left(x,\tau \right)
\label{S-Ham}
\end{equation}
admits the additional invariance
\begin{equation}
\psi (x,\tau ) \rightarrow
\exp \left[ \dfrac{ie}{\hbar c}\Lambda (x,\tau )\right] \, \psi (x,\tau)
\label{wave}
\end{equation}
when taken together with (\ref{gauge_1}). 

The generalization of Stueckelberg's framework to include $\tau$-dependent
fields and gauge transformations thus succeeds in implementing his model of pair
processes in classical mechanics, but~also raises new questions.  Perhaps
most significantly, while defining the system in an unconstrained 8D phase space
relaxes the a priori mass shell relation $\dot x^2 = c^2 $ and thus
permits classical trajectories that reverse the direction of their time
evolution, it also eliminates reparameterization invariance. {{This is
because the mass
shell constraint and reparameterization invariance are related features of a Lagrangian
that is homogeneous of first degree in the velocities, which is not the case for}
(\ref{lagrangian})}. %%% footnote moved to paragraph.
 Moreover, in~Stueckelberg--Horwitz--Piron (SHP) electrodynamics, the
evolution parameter $\tau$ cannot be identified as the proper time of the
motion, but is a dynamical quantity proportional to it through 
\begin{equation}
c^2 \; ds^2 (\tau) = -g_{\mu\nu} dx^\mu dx^\nu = -\dot x^2 (\tau) \ d\tau^2 \qquad 
g_{\mu\nu} = \text{diag}(-1,1,1,1) .
\end{equation}

Therefore, the parameter $\tau$ plays the role of an irreducible chronological
time, independent of the~spacetime coordinates and similar to the external time
$t$ in nonrelativistic Newtonian mechanics.  It determines the temporal ordering
of events, the order of their physical occurrence, which may be~different
from the order of observed coordinate times $x^0$ registered by laboratory
clocks as the~events appear in measuring apparatus.  For example, in a
classical, continuous version of a Feynman spacetime diagram, the event
trajectory  
\begin{equation}
x(\tau) = \left( c t (\tau) , {\mathbf x}(\tau) \right) 
= \left( c \left( t_0 + \dot t_0 \tau - \tfrac12 \ddot t_0 \tau^2 \right) , {\mathbf x}(\tau) \right) 
\end{equation}
where $t_0$, $\dot t_0$ and $\ddot t_0$ are positive constants, reverses time direction
at $\tau_0 = \dot t_0 / \ddot t_0$, and represents a pair annihilation process.
Singling out three events in their $\tau$-chronological order of occurrence,
\begin{equation}
\begin{array}{ccc}
\tau =0 &x^0 = ct_0& \dot x^0 = c\dot t_0 \strt{12} \\
\tau =\tau_0 \qquad &x^0 = c \left( t_0 + \dot t_0^2 / 2\ddot t_0\right)
\qquad  & \dot x^0 = 0 \strt{12} \\
\tau =2\tau_0 &x^0 = c t_0&\dot x^0 = -c\dot t_0  \strt{10}
\end{array}
\end{equation}
the laboratory apparatus will first record both particle and antiparticle
trajectories at $x^0 = ct_0$, then~the~annihilation event at
$x^0 = c \left( t_0 + \dot t_0^2 / 2\ddot t_0\right)$
and no particles for
subsequent values of $x^0$.  The classical antiparticle is identified here by
its negative energy and no charge conjugation operation is
necessary.  {{Looking ahead to the current defined in} (\ref{curr-def}), {one
sees that the electric charge  $Q = \int d^3x \; j^0 (x,\tau)$ will similarly
reverse sign on this section of the trajectory.}}  %%% footnote moved to paragraph
Although the laboratory
apparatus observes the two events at $x^0 = c t_0$ as simultaneous, they may be
distinguished in SHP theory by other $\tau$-dependent interactions.  

As Horwitz has observed, grandfather paradoxes may be resolved by noticing that
the return trip to a past coordinate time $x^0$ must take place while the
chronological time $\tau$ continues to increase.  The~{\em {occurrence}} of event
$x^\mu(\tau_1)$ at $\tau_1$ is understood to be an irreversible process that
cannot be~changed by a subsequent event occurring at the same spacetime
location, $x^\mu(\tau_2) = x^\mu(\tau_1)$ with $\tau_2 > \tau_1$.  This~absence
of closed time-like curves similarly applies in SHP quantum 
electrodynamics~\cite{qft} where the particle propagator $G(x_2 - x_1,\tau_2 - \tau_1)$ vanishes
unless $\tau_2 > \tau_1$, thus preventing divergent matter loops, when $x_2 =
x_1$. {This } $\tau$-{retarded causality was first shown by Feynman to be
equivalent to the Feynman 
contour for propagators in connection with the
path integral for the Klein--Gordon equation}  \cite{Feynman-1,Feynman-2}. 
{In SHP quantum electrodynamics (QED), it also
emerges from the vacuum expectation value of $\tau$-ordered operator products}.
%%% confirm the definition for QED
In the microscopic event dynamics described by this explicit distinction between
chronological and coordinate time \cite{Two-Aspects}, a covariant Hamiltonian
generates evolution of a 4D block universe defined at $\tau$ to an
infinitesimally close 4D block universe defined at \hbox{$\tau + d\tau$}.  
Standard Maxwell electrodynamics emerges as an equilibrium limit in which the
system becomes \hbox{$\tau$-independent}, and~the~4D block universe remains
static.  The details of this \hbox{$\tau$-dependence} in~the~interacting fields and
currents can be studied by reconciling classical SHP with
classical Maxwell electromagnetic~phenomenology.  

\section{Classical SHP Electrodynamics}

In analogy with the notation $x^0 = ct$, we adopt the formal designations
\begin{equation}
x^5 = c_5 \tau \qquad   \partial_5 = \dfrac{1}{c_5} \partial_\tau \quad  
\quad j^5 = c_5 \rho \qquad  a_5 =  \dfrac{1}{c_5} \phi
\end{equation}
and the conventions 
\begin{equation}
\mu ,\nu =0,1,2,3 \qquad \alpha ,\beta ,\gamma =0,1,2,3,5 \qquad
g_{\alpha\beta} = \text{diag}(-1,1,1,1,\pm 1)
\end{equation}
so that $c_5 / c$ characterizes the relative rate of evolution in $\tau$, and as
shown in Section \ref{Sec.4}, SHP becomes Maxwell theory in
the limit $c_5 \rightarrow 0$. 
Writing (\ref{lagrangian}) and (\ref{f-def}) as
\begin{equation}
L=\dfrac{1}{2}M\dot{x}^{\mu }\dot{x}_{\mu }+\dfrac{e}{c}\dot{x}^{\alpha }a_{\alpha }
\qquad 
f^\alpha_{\; \;  \beta} = \partial^\alpha a_\beta - \partial_\beta a^\alpha
\end{equation}
the event dynamics are given by the Lorentz force
\begin{equation}
M\ddot{x}^{\mu } = \dfrac{e}{c} f^\mu_{\; \;  \alpha} (x,\tau )\dot{x}^\alpha
\qquad 
\frac{d}{d\tau }(-\tfrac{1}{2}M\dot{x}^{2})= g_{55} \dfrac{e c_5}{c}\; f^{5\mu }\dot{x }_{\mu}
\label{lorentz}
\end{equation}
equivalent to substituting the $\tau$-dependent fields (\ref{fields}) into
Stueckelberg's evolution Equation (\ref{10}) in flat spacetime.  For the moment, we 
understand the factor $g_{55}$ as a choice of sign for the second of
(\ref{lorentz}) and not a hint at some 5D metric structure.

To complete the dynamical picture, we re-express the velocity-potential
interaction as a~current-potential integral  
\begin{equation}
\begin{array}{c}
\dot{X}^{\alpha } a_{\alpha } \rightarrow \dint d^{4}x
\; \dot{X}^\alpha(\tau) \delta^4\left(x-X(\tau)\strt{4} \right) a_{\alpha }(x,\tau ) =
\dfrac{1}{c}\dint d^{4}x \; j^\alpha (x,\tau) a_\alpha (x,\tau) \strt{16} \\ 
j^\alpha (x,\tau) = c\dot{X}^\alpha(\tau) \delta^4\left(x-X(\tau)\strt{4} \right)
\end{array}
\label{curr-def}
\end{equation}
and choose some kinetic action term for the fields, the most obvious candidate
being 
\begin{equation}
{\cal L}_\text{kinetic} =
\frac{1}{4c} f^{\alpha \beta }(x,\tau )f_{\alpha \beta }\left( x,\tau\right)
\label{kin-0}
\end{equation}
{which generalizes to 5D the standard Maxwell action \cite{jackson}.
Combining (\ref{curr-def}) and (\ref{kin-0}), the electromagnetic~action}
\begin{equation}
S_\text{em} = \int d^{4}xd\tau \left\{\dfrac{e}{c^2} \; j^{\alpha }(x,\tau )a_{\alpha }(x,\tau )-
\frac{1}{4c} f^{\alpha \beta }(x,\tau )f_{\alpha \beta }\left( x,\tau\right)
\right\}
\label{action-0}
\end{equation}
can be varied with respect to $a_\alpha$ to produce Maxwell-like field
equations, admitting a wave equation with associated Green's function whose
solutions describe the fields induced by specific event trajectories.  However,
one is immediately confronted by conceptual difficulties in attempting to
describe even the simple case of low energy Coulomb scattering.  First, although
the current $j^\alpha(x,\tau )$ and the potential $a_\alpha(x,\tau )$ are
individually constructed to be vector $+$ scalar representations of O(3,1) on
physical grounds, the 5D scalar structure of the electromagnetic action
(\ref{action-0}) places all five components on the same algebraic footing.  This
suggests an underlying formal symmetry larger than O(3,1) but containing it as a
subgroup: O(4,1) for the choice $g_{55} = 1$ or O(3,2) symmetry for $g_{55} =
-1$.  The~formal 5D scalar structure survives in the wave equation and these
symmetry considerations cannot be~entirely ignored.  Second, because $x^\mu$ and
$\tau$ are introduced to play very different roles, the rest frame of an event
is $\tau$-dependent.  Thus, a ``static'' particle---an event evolving uniformly
along the $x^0$ axis in its rest frame---is described by $x_{\text{rest}} =
\left( c\tau,{\mathbf 0} \right) $.  It turns out that the potential at an
observation point $x = (ct,{\mathbf x})$ induced by this particle is of the form
$\delta\left( (\tau - t)^2 - \vert{\mathbf x}\vert^2 / c^2\right) $ with support
sharply focused on the lightcone of the source event's immediate location.  A
test event evolving uniformly as $x_{\text{test}} = \left( c(\tau +
t_0),{\mathbf R} \right) $ will experience the potential $\delta\left(  t_0^2 -
\vert{\mathbf R}\vert^2 / c^2\right) $ for all $\tau$, rendering comparison with
experiment nearly impossible.  It is worth noting that the problem of clock
synchronization between the~source and test events (characterized here by $x^0_{\text{test}}
(0) = ct_0$) is~not a problem for the~quantized theory where the use of asymptotic
states with sharp mass implies maximum uncertainty in the~location $x^\mu$ of the event
in $\tau$.  By introducing a degree of uncertainty 
into the definition of the~classical electromagnetic
action, a reasonable theory may be constructed.

\section{Non-Local Field Kinetics $\ \leftrightarrow \ $ Ensemble of Events} 

The problem of potentials with $\delta$-function support can be repaired by writing the action
with a~slightly less obvious candidate for the field kinetic term, the non-local form
\begin{equation}
S_\text{em} = \int d^{4}xd\tau \left\{\dfrac{e}{c^2} \; j^{\alpha }(x,\tau )a_{\alpha }(x,\tau )-
\int \frac{ds}{\lambda}\, \frac{1}{4c} \left[f^{\alpha \beta }(x,\tau )\Phi (\tau -s)f_{\alpha \beta
}\left( x,s\right) \right] \right\}
\label{action}
\end{equation}
where $\lambda$ is a parameter with dimensions of time.  The field interaction kernel is
\begin{equation}
\Phi (\tau ) = \delta \left( \tau \right) -(\xi\lambda) ^{2}\delta ^{\prime \prime
}\left( \tau \right) = \int \frac{d\kappa}{2\pi} \,\left[ 1+\left( \xi\lambda
\kappa \right) ^{2}\right] \,e^{-i\kappa \tau } 
\label{kernel}
\end{equation}
where
\begin{equation}
\xi =\frac{1}{2}\left[ 1+\left( \frac{c_5}{c}\right) ^{2}\right]
\label{xi}
\end{equation}
is chosen so that the low energy Lorentz force agrees with Coulomb's law. 
The kinetic term now includes
$\left( \partial_\tau f^{\alpha \beta }(x,\tau )\right) \left( \partial_\tau
f_{\alpha \beta }\left( x,\tau\right)\right) $
which explicitly breaks the 5D symmetry to O(3,1) while maintaining gauge
symmetry.  We write the inverse function of the interaction kernel as
\begin{equation}
\varphi (\tau ) = \lambda \Phi^{-1}  (\tau )= \lambda \int \frac{d\kappa}{2\pi}
\,\frac{e^{-i\kappa \tau }}{
1+\left(\xi  \lambda \kappa \right) ^{2}}
=\frac{1}{2\xi}e^{-\vert\tau \vert/\xi \lambda }
\label{inv}
\end{equation}
which satisfies
\begin{equation}
\int \frac{ds}{\lambda} ~\varphi \left( \tau - s \right)  \Phi  \left( s \right)  = \delta (\tau)
\qquad
\int \frac{d\tau}{\lambda} ~\varphi  \left( \tau \right)  =1 .
\label{inv-2}
\end{equation}

Varying the action (\ref{action}) with respect to the potentials,
leads to field equations 
\begin{equation}
\partial _{\beta }f^{\alpha \beta }_\Phi (x,\tau) =
\partial _{\beta }\int \frac{ds}{\lambda}\,\Phi (\tau -s)f^{\alpha \beta }(x,s)
=\frac{e}{c}\ j^{\alpha }(x,\tau )
\end{equation}
describing the non-local superposition of fields $f^{\alpha \beta }_\Phi$ sourced by the
instantaneous event current $j^\alpha(x,\tau )$.  Using (\ref{inv-2}) to remove
$\Phi(\tau)$ from the LHS and writing the
Bianchi identity, we obtain equations for the~local field sourced by a non-local
superposition of event currents,  
\begin{eqnarray}
&&\partial _{\beta }f^{\alpha \beta }\left( x,\tau \right)
= \frac{e}{c}\int ds~\varphi \left( \tau -s\right)
j^{\alpha }\left( x,s\right)
= \dfrac{e}{c} \, j_\varphi^{\alpha } \left( x,\tau \right) \strt{22}
\label{gauss}\\
&&\partial _{\alpha }f_{\beta \gamma } + 
\partial _{\gamma }f_{\alpha \beta } + 
\partial _{\beta }f_{\gamma \alpha }  = 0
\label{pm-h}
\end{eqnarray}
which are formally similar to Maxwell's equations in 5D and are called
pre-Maxwell equations.
Rewriting the field equations in 4D tensor, vector and scalar components, they take the form
\begin{equation}
\begin{array}{lcl}
\partial _{\nu }\;f^{\mu \nu }- \dfrac{1}{c_5} \dfrac{\partial}{\partial \tau}\;f^{5\mu
}=\dfrac{e}{c} \; j^{\mu }_\varphi 
& \mbox{\qquad} &\partial _{\mu }\;f^{5\mu }=\dfrac{e}{c}\;  j^5_\varphi
= \dfrac{c_5}{c} \; e \rho_\varphi
\vspace{8pt}\\ 
\partial _{\mu }f_{\nu \sigma }+\partial _{\nu }f_{\sigma \mu }+\partial _{\sigma }f_{\mu \nu }=0 
& &\partial _{\nu }f_{5\mu }-\partial _{\mu }f_{5\nu }
+ \dfrac{1}{c_5} \dfrac{\partial}{\partial \tau}f_{\mu \nu }=0 \\
\end{array}
\label{premax}
\vspace{4pt}
\end{equation}
which may be compared with the 3-vector form of Maxwell's equations
\begin{equation}
\vspace{4pt}
\begin{array}{lcl}
\nabla \times \mathbf{B}-\dfrac{1}{c} \dfrac{\partial}{\partial t}\mathbf{E}=\dfrac{e}{c} \ \mathbf{J} 
& \mbox{\qquad} \mbox{\qquad} & \nabla \cdot \mathbf{E}=\dfrac{e}{c} \ J^{0} \vspace{8pt}\\ 
\nabla \cdot \mathbf{B}=0
& &\nabla \times \mathbf{E}+\dfrac{1}{c} \dfrac{\partial}{\partial t}\mathbf{B}=0 \\
\end{array}
\end{equation}
showing that $f^{5\mu }$ plays the role of the vector electric field and $f^{\mu \nu
}$ plays the role of the magnetic field.  It follows
from (\ref{gauss}) that current conservation takes the form
\begin{equation}
\partial_\alpha j^\alpha = \partial_\mu j^\mu \left( x,\tau \right) +
\frac{1}{c_5} \partial_\tau j^5 \left( x,\tau \right) = 0
\end{equation}
so that a change in the divergence in the 4D Maxwell-like current $j^\mu$ must
be compensated by the~addition or subtraction of events through $j^5$.
The fifth component of the current 
$j^5\left( x,\tau \right) = c_5 \rho \left( x,\tau \right)$ thus
plays the role of an event density, the probability that a material event occurs
at the spacetime point $x$ at the chronological time $\tau$.  Integrated over
all spacetime,
\begin{equation}
\frac{d}{d\tau} \int d^4 x \ \rho \left( x,\tau \right)
= - \int d^4 x \ \partial_\mu j^\mu \left( x,\tau \right) = 0 
\end{equation}
expresses the conservation of total event number.

Rewriting the source of the inhomogeneous Equation (\ref{gauss}) as
\begin{equation}
j_\varphi^{\alpha } \left( x,\tau \right)
= \int ds~\varphi \left( \tau -s\right) j^{\alpha }\left( x,s\right)
= \frac{1}{2\xi}\int ds~e^{-\vert s \vert/\xi \lambda } \ j^{\alpha }\left( x,\tau -s\right)
\label{phi-cur}
\end{equation}
we recognize $j_\varphi^{\alpha } \left( x,\tau \right)$ as a weighted
superposition of currents, each originating at an event $X^\mu(\tau - s)$
displaced from $X^\mu(\tau)$ by an amount $s$ along the worldline.  It is useful
to regard this superposition as the current produced by an ensemble of events in
the neighborhood of $X^\mu(\tau)$, a~view encouraged by the particular weight
function $\varphi (s)$.  Given a Poisson distribution describing the~occurrence
of independent random events with a constant average rate of $1 / \lambda \xi$
events per second, the~average time between events is $\lambda \xi$ and the
probability  at $\tau$ that the next event will occur following a time interval
$s > 0$ is just \hbox{$\varphi (s) / \lambda = e^{- s /\xi \lambda } / \xi
\lambda$}.  Extending the displacement to positive and negative values, the
ensemble is constructed by assembling a set of instantaneous event currents
$j^\alpha \left( x,\tau -s\right)$ along the worldline, each weighted by
$\varphi (s)$, the probability that the occurrence of this event is delayed from
$\tau$ by an interval of at least $\vert s \vert$.  We will see that the leading
term of the Green's function manifestly breaks 5D symmetry to O(3,1) and the
causality relations embedded in this term select the one event from this
ensemble for which the interacting events are at light-like separation, depending
on their relative $\tau$-synchronization.  

The ensemble can also be understood by thinking of
$j_\varphi^{\alpha } \left( x,\tau \right)$ as a random variable
describing the~probability of finding a current density at $x$ at a given
$\tau$.  Then we may consider a correlation function for the event density of the type 
\begin{equation}
\left\langle \rho  \left( \tau \right) \strt{4} \rho \left( s \right) \right\rangle = 
\frac{1}{\cal{N}} \int d^4 x \ \rho 
\left( x, \tau \right) \rho \left( x, s \right) 
\end{equation}
where $\cal{N}$ is a normalization.
For a uniformly moving event with \hbox{$X^\mu (\tau) = u^\mu \tau$}, the raw event
current~(\ref{curr-def}) leads to
\begin{equation}
\left\langle \rho  \left( \tau \right) \strt{4} \rho \left( s \right) \right\rangle
= \frac{c^2}{\cal{N}} \int d^4 x \ \delta^4 \left( x - u \tau \right)
\delta^4 \left( x - u s \right)
= \frac{c^2\delta^3 \left( {\mathbf 0} \right)}{\vert u^0 \vert \cal{N} } \ \delta (\tau - s)
\end{equation}
showing that the currents at differing times $\tau \ne s$ are uncorrelated. 
For the ensemble current defined in (\ref{gauss}), the correlation becomes
\begin{eqnarray}
\left\langle \rho_\varphi  \left( \tau \right) \strt{4}
\rho_\varphi \left( s \right) \right\rangle
\eq \frac{c^2}{\cal{N}}  \int d\tau^\prime ds^\prime d^4 x \ \varphi(\tau - \tau^\prime)
\varphi(s - s^\prime)
\delta^4 \left( x - u \tau^\prime \right)
\delta^4 \left( x - u s^\prime \right) \notag \\
\eq \frac{c^2 \delta^3 \left( {\mathbf 0} \right)}{\vert u^0 \vert \cal{N}} \
\int d\tau^\prime \ \varphi(\tau - \tau^\prime) \varphi(\tau^\prime - s) \notag \\
\eq \frac{c^2 \delta^3 \left( {\mathbf 0} \right)}{4 \xi^2 \vert u^0 \vert \cal{N}} \
\int d\tau^\prime \ e^{-\vert\tau - \tau^\prime \vert/\xi \lambda 
-\vert\tau^\prime - s \vert/\xi \lambda }
\end{eqnarray}
so that taking $\tau > s$ and separating the integral into three intervals
punctuated by $s$ and $\tau$ 
leads to
\begin{equation}
\left\langle \rho_\varphi  \left( \tau \right) \strt{4}
\rho_\varphi \left( s \right) \right\rangle
= \frac{\lambda c^2 \delta^3 \left( {\mathbf 0} \right)}{4 \xi \vert u^0 \vert \cal{N}} \
\left( 1 + \frac{\tau - s}{ \xi \lambda } \right) 
e^{-\left( \tau - s \right) /\xi \lambda} 
\end{equation}
with a time dependence characteristic of an Ornstein--Uhlenbeck process.
Regarding the smoothed current $j^\alpha_\varphi (x,\tau)$ 
produced by an event $X^\mu (\tau)$ as the
instantaneous current produced by an ensemble of events, this 
correlation suggests that the ensemble is the result of the Brownian motion
found by subjecting $X^\mu (\tau)$ to a random force under viscous drag. 

The pre-Maxwell equations in Lorenz gauge lead to the wave equation 
\begin{equation}
\partial _{\beta }\partial ^{\beta }a^{\alpha }=(\partial _{\mu }\partial
^{\mu }+\partial _{\tau }\partial ^{\tau })a^{\alpha }=(\partial _{\mu
}\partial ^{\mu } +  \frac{g_{55}}{c_5^{2}} \; \partial _{\tau }^{2})a^{\alpha }=-
\frac{e}{c}\ j_{\varphi }^{\alpha }\left( x,\tau \right)
\end{equation}
whose solutions may respect 5D symmetries broken by the O(3,1) symmetry of the event
dynamics.  A~Green's function solution to
\begin{equation}
(\partial_{\mu}\partial^{\mu} + \frac{g_{55}}{c_5^{2}} \; \partial _{\tau }^{2})G(x,\tau)
=- \delta^4\left( x \right) \delta \left( \tau \right)
\end{equation}
can be used to obtain potentials in the form
\begin{equation}
a^{\alpha }\left( x,\tau \right) = -\frac{e}{c} \int d^{4}x^{\prime }d\tau
^{\prime } \ G\left( x-x^{\prime }, \tau -\tau ^{\prime }\right) j_{\varphi
}^{\alpha }\left( x^{\prime },\tau^{\prime } \right) .
\label{green-pot}
\end{equation}
The principal part Green's function \cite{green} is  
\begin{eqnarray}
G_{P}(x,\tau ) \eq -{\frac{1}{{2\pi }}}\delta (x^{2})\delta (\tau )-{\frac{c_5}{{
2\pi ^{2}}}}{\frac{\partial }{{\partial {x^{2}}}}}{\theta (-g_{55}g_{\alpha
\beta }x^{\alpha }x^{\beta })}{\frac{1}{\sqrt{-g_{55}g_{\alpha \beta
}x^{\alpha }x^{\beta }}}}
\label{greens} 
\\  \eq \  G_{Maxwell} + G_{Correlation} 
\end{eqnarray}
where $G_{Maxwell}$ breaks a higher symmetry to O(3,1) while the support of
$G_{Correlation}$ is  
\begin{equation}
{-g_{55}g_{\alpha \beta }x^{\alpha }x^{\beta }=}\left\{ 
\begin{array}{lll}
-\left( x^{2}+c_{5}^{2}\tau ^{2}\right) =c^{2}t^{2}-\mathbf{x}%
^{2}-c_{5}^{2}\tau ^{2}>0 & , & {g_{55}=1} \strt{16}\\ 
\left( x^{2}-c_{5}^{2}\tau ^{2}\right) =\mathbf{x}^{2}-c^{2}t^{2}-c_{5}^{2}%
\tau ^{2}>0 & , & {g_{55}=-1}%
\end{array}%
\right.
\label{gf}
\end{equation}
with causality properties dependent on the choice of $g_{55}$.

The contribution from $G_{Correlation}$ is smaller than that of $G_{Maxwell}$
by $c_5 / c$ and drops off as
$1/\left\vert \mathbf{x} \right\vert^2$, so it may be neglected at low energy \cite{speeds}.
The contribution to the potential from $G_{Maxwell}$ is
\begin{eqnarray}
a^{\alpha }\left( x,\tau \right) &&\mbox{\hspace{-20pt}}=-\frac{e}{c} \int
d^{4}x^{\prime }d\tau ^{\prime } \; G_{Maxwell}\left( x-x^{\prime } , \tau -\tau
^{\prime }\right) j_{\varphi }^{\alpha }\left( x^{\prime },\tau^{\prime }
\right) 
\notag \\
&&\mbox{\hspace{-20pt}}=\frac{e}{2\pi c} \int
d^{4}x^{\prime }d\tau ^{\prime } ds \;
\delta \left[ \left( x-x^\prime \right)^2\right] 
\delta (\tau -\tau^\prime ) \; \varphi \left( \tau^\prime -s\right)
j^{\alpha }\left( x^{\prime },s \right)
\end{eqnarray}
and because $G_{Maxwell}$ has support at equal-$\tau$ this can be written
\begin{equation}
a^{\alpha }\left( x,\tau \right) = \frac{e}{2\pi c} \int ds \; \varphi \left( \tau -s\right)
\int d^{4}x^{\prime } \;
\delta \left[ \left( x-x^\prime \right)^2\right] 
\; j^{\alpha }\left( x^{\prime },s \right)
\end{equation}
expressing the potential as an ensemble of single-event potentials.  
Inserting the current defined in (\ref{curr-def}) 
\begin{equation*}
j^{\alpha }\left( x,\tau \right) = c \dot X^\alpha (\tau) \delta^4 (x-X(\tau))
\end{equation*}
and using the identity
\begin{equation}
\dint d\tau f\left( \tau \right) \delta \left[g\left( \tau \right) \right]
=\dfrac{f\left( \tau_R\right) }{\left\vert g^{\prime }\left( \tau_R\right)
\right\vert } ,
\label{identity}
\end{equation}
where $\tau_R$ is the retarded time that solves
\begin{equation}
g\left( \tau \right) =(x-X(\tau_R))^2 =0 \qquad
\theta ^{ret}=\theta \left( x^{0}-X^{0}\left( \tau_R\right) \right) ,
\end{equation}
we find the potential
\begin{eqnarray}
a^{\alpha }\left( x,\tau \right) \eq \frac{e}{2\pi } \int ds \; \varphi \left( \tau -s\right)
 \dot X^\alpha (s) \;
\delta \left[ \left( x-X^\alpha (s)  \right)^2\right] 
\notag \\
\eq \frac{e}{4\pi }\varphi \left( \tau -\tau_R\right) \frac{
\dot{X}^{\alpha }\left( \tau_R\right) }{\left\vert \left( x^{\mu }-X^{\mu }\left( \tau_R\right)
\right) \dot{X}_{\mu }\left( \tau_R\right) \right\vert}
\end{eqnarray}
which is the standard Li\'{e}nard--Wiechert potential multiplied by $\varphi
\left( \tau -\tau_R\right)$.  Thus, while the current that sources the
pre-Maxwell field
represents an ensemble of events along the worldline, the retarded causality of
the Green's function selects the one member of the ensemble that intersects the
lightcone of the observation point.  The remaining \hbox{$\tau$-dependence} of the
fields resides in the finite function $\varphi$ and expresses the relative
time synchronization between the source and a test event experiencing 
the~potential at the spacetime point $x$ at the chronological time $\tau$.

To find the Coulomb potential, we specify the event trajectory \hbox{$X\left(
\tau \right) =\left(c \tau ,\mathbf{0}\right)$} which produces the~instantaneous current 
\begin{equation}
j^0(x,\tau) = c^2 \delta(t-\tau) \, \delta^3({\mathbf x}) \qquad {\mathbf
j}(x,\tau) = 0  \qquad j^5 (x,\tau) = cc_5 \delta(t-\tau) \, \delta^3({\mathbf x})
\end{equation}
and so the potential
\begin{equation}
a^{0} (x,\tau) =  {\frac{e}{{4\pi \vert \mathbf{x} \vert }}}\varphi
\left(\tau - \left( t - \frac{\vert \mathbf{x} \vert}{c} \right) \right) \qquad \mathbf{a} = 0 
\qquad a^{5}(x,\tau) = \frac{c_5}{c} a^{0} (x,\tau)
\end{equation}
is found from from $G_{Maxwell}$.
Consider a test event \hbox{$x(\tau )=\left( c(\tau - \tau_0) ,\mathbf{x}\right) $}
evolving along a parallel trajectory at spacial separation $\mathbf{x}$ and time
offset $\tau_0$ so that
\begin{equation}
\varphi\left(\tau - \left( t - \frac{\vert \mathbf{x} \vert}{c} \right) \right)
= \frac{1}{2\xi } e^{-\frac{1}{\xi \lambda }\left\vert \tau_0 + \left\vert
\mathbf{x}\right\vert / c \strt{4} \right\vert } . 
\end{equation}
If the test event is precisely on the forward lightcone of the source with $\tau_0 =
-\vert \mathbf{x}\vert / c$ then 
\hbox{$\varphi = 1 $} and the interaction is purely Coulomb in form.  If $\tau_0 = 0$
so that the events are synchronized at $x^0$, then
\begin{equation}
a^{0}(x,\tau )  
=\frac{1}{2\xi }
{\frac{e}{{4\pi \vert \mathbf{x} \vert }}}
e^{- \left\vert \mathbf{x}\right\vert  /\xi \lambda c}
\label{yukawa}
\end{equation}
which has the form of a Yukawa-type potential 
with photon mass $m_\gamma \sim \hbar / \xi \lambda c^2 $.
If the source and test events are slightly desynchronized, with $\tau_0 > 0$, then 
the interaction is weakened by a factor $e^{-\tau_0  /\xi \lambda c }$.  

Thus, the factor $\lambda$ that characterizes the width of the ensemble and
represents the average
event inter-occurrence time, also determines the mass spectrum of the photons mediating
the interaction between events.  As proposed by Stueckelberg and seen in the
second of (\ref{lorentz}), the masses of the~particles and fields are not
separately conserved, although Noether's theorem for affine $\tau$-displacement
symmetry guarantees that the total mass is a constant of the motion
\cite{lorentz}.  If $\lambda$ is small (so that $\varphi / \lambda$ approaches a
delta function and the current narrows to a small neighborhood around the event),
the~mass spectrum becomes wide and the interaction range and cross-section
decreases.  If $\lambda$ is large, the~support of the
current spreads along the worldline, the~potential becomes Coulomb-like and 
the~photon mass spectrum is small.  

A similar role is seen for $\lambda$ in SHP quantum field theory.  From the Fourier
expansion for the~electromagnetic Green's function (\ref{greens})
\begin{equation}
G(x-x',\tau-\tau') = \int \dfrac{d^4 k d\kappa}{(2\pi)^5} \;
\frac{e^{i[ k\cdot (x-x') + g_{55} \kappa c_5(\tau-\tau')]}}
{k^2 +g_{55} \kappa^2 -i\epsilon}
\label{eqn:5.48}
\end{equation}
it appears that photon loops in the 5D theory would render it non-renormalizable.  However,
quantization of the higher order field kinetic term leads to the photon propagator factor
\begin{equation}
\left[ g^{\mu \nu }-\frac{k^{\mu }k^{\nu }}{k^{2}}\right] 
\frac{-i}{k^{2}+g_{55} \kappa ^{2}-i\epsilon }\ \ \frac{1}{1+\left( \lambda
\xi\right) ^{2}\kappa ^{2}
}  
\label{phot-prop}
\end{equation}
and the theory is super-renormalizable at second order.  Again, we notice that if
$\lambda$ is large, then large values of photon mass $\kappa$ are suppressed.

While mass exchange must be present in any classical theory of pair
processes and must also be small to account for standard electromagnetic
phenomenology, such a compromise cannot explain the fixed masses of elementary
particles.  However, it has been shown that under certain circumstances
\cite{mass} a self-interaction induced through $G_{Correlation}$ has the effect
of restoring on-shell evolution in event trajectories and thus returning the
particle worldline to the observed fixed mass.  As~seen in (\ref{gf}) when
$g_{55} = 1$, the Green's function has time-like support, permitting the event to
interact with the field produced earlier along its worldline.  The net effect of
this self-interaction is a damping (or anti-damping) force that accelerates the
event evolution to its asymptotic mass shell.
A more general approach is found
in the statistical mechanics of the many-event system. 
While the model presented here describes a particle
as a weighted ensemble of events $\varphi(s) X^\mu(\tau - s)$ along a single
worldline, Horwitz has modeled \cite{chemical} a particle as an ensemble of
$n$ independent spacetime events $X^\mu_i(\tau)$, $i = 1, 2, \ldots, n$ defined at a
given $\tau$.  He has shown that the total particle mass is determined by a
chemical potential.  Following collisions governed by a general class of
interactions that includes pair processes, particles return to their equilibrium
mass values.  These developments indicate that the~statistical mechanics of
event ensembles in the construction of classical particles will be a fruitful
way to understand mass and perhaps derive masses from first principles.

\section{Maxwell Theory as an Equilibrium State of SHP \label{Sec.4}}

{Following an argument by Stueckelberg, Saad et al. \cite{saad} noticed that under the
boundary~conditions }
\begin{equation}
j^5_\varphi (x,\tau) \underset{\tau \rightarrow \pm \infty }{\mbox{\
}\xrightarrow{\hspace*{1.5cm}}\mbox{\ }} 0 \qquad 
f^{5\mu} (x,\tau) \underset{\tau \rightarrow \pm \infty }{\mbox{\
}\xrightarrow{\hspace*{1.5cm}}\mbox{\ }} 0
\end{equation}
integration of the pre-Maxwell equations provide
\begin{equation}
\left. 
\begin{array}{c}
\partial _{\beta }f^{\alpha \beta }\left( x,\tau \right)
=\dfrac{e}{c} j_{\varphi}^{\alpha }\left( x,\tau \right) \\ 
\\ 
\partial _{\lbrack \alpha }f_{\beta \gamma ]}=0 \\
\\ 
\partial _{\alpha }j^{\alpha } = 0
\end{array}
\right\}
\underset{\dint \dfrac{d\tau}{\lambda} }{\mbox{\quad}\xrightarrow{\hspace*{1.8cm}}
\mbox{\quad}}\left\{ 
\begin{array}{c}
\partial _{\nu }F^{\mu \nu }\left( x\right) =\dfrac{e}{c}J^{\mu }\left( x\right) 
\\
\\
\partial _{\nu }F^{5 \nu }\left( x\right) =\dfrac{e}{c}J^{5 }\left( x\right)
\\ 
\\ 
\partial _{\lbrack \mu }F_{\nu \rho ]}=0 \\
\\ 
\partial _{\mu } J^{\mu}(x) = 0
\end{array}
\right.
\end{equation}
where
\begin{equation}
A^{\alpha }(x)=\int \frac{d\tau}{\lambda}  \;a^{\alpha }(x,\tau )
\qquad F^{\alpha \nu }(x)=\int \frac{d\tau}{\lambda}  \;f^{\alpha \nu }(x,\tau )
 .
\end{equation}
This integration has been called concatenation and is understood as aggregation
of all events that occur at a spacetime point $x$ over all $\tau$.  The
decoupling of $F^{5\mu}$ from $F^{\mu\nu}$, which now satisfies Maxwell's
equations, suggests that the SHP can be seen as an underlying microscopic dynamics for
which Maxwell theory is an equilibrium or expectation state.  In particular, we
see that integration of~(\ref{yukawa}) provides the Coulomb potential and
using (\ref{inv-2})
to integrate the 4-vector current yields the~Maxwell current $J^\mu(x)$ in the
standard form as
\begin{eqnarray}
\int d\tau \; j_\varphi^{\mu } \left( x,\tau \right)
\eq c \int  d\tau \;ds~\varphi \left( \tau -s\right) \dot{X}^\mu(s)
\delta^4\left(x-X(s)\strt{4} \right) \notag \\
\eq c \int  d\tau \; \dot{X}^\mu(\tau)
\delta^4\left(x-X(\tau)\strt{4} \right) .
\label{int-cur}
\end{eqnarray}
Similarly, concatenation of (\ref{greens}) 
\begin{equation}
\int d\tau \ G_{Maxwell} = D(x) =  -{\frac{1}{{2\pi }}}\delta (x^{2})
\qquad 
\int d\tau \ G_{Correlation} = 0 
\end{equation}
recovers the 4D Maxwell Green's function.  

Another approach \cite{speeds} to retrieving Maxwell theory from SHP is to slow
the $\tau$-evolution to zero by taking $c_5 / c \rightarrow 0$, thus freezing
the microscopic system into a static equilibrium.  Under this condition, 
the homogeneous Equation (\ref{pm-h}) imposes the condition 
\begin{equation}
c_5 \left( \partial _{\nu }f_{5\mu }-\partial _{\mu }f_{5\nu } \right) +
\partial _{\tau }f_{\mu \nu }=0\underset{ c_5 \rightarrow 0
}{\mbox{\ \  }\xrightarrow{\hspace*{1.5cm}} \mbox{\ \  }} \partial _{\tau }f_{\mu
\nu } = 0 
\label{stat-f}
\end{equation}
requiring that the field strength $f^{\mu\nu}$ be $\tau$-independent in this
limit.  As seen in the Li\'{e}nard--Wiechert potential, the
$\tau$-dependence resides in $\varphi (\tau - \tau_R)$ and can only be 
suppressed by taking $\lambda \rightarrow \infty$.  
We recall that $\lambda$ is the correlation time for the current
$j^\alpha$, that is the measure of information about the interaction at time
$\tau$ available from observation of the interaction at time $s < \tau$.
Thus, the action-at-a-distance Coulomb law can be understood as the effect of 
a long-term correlation for the instantaneous potential.
Applying the combined limits on $c_5 / c$ and $\lambda$ to (\ref{xi}) and (\ref{inv}),
we find $ \varphi (\tau ) \rightarrow 1/2\xi = 1$ and so all field 
components must be $\tau$-independent.  This requirement effectively
assigns equal weight to all event
currents $j^\alpha(x,\tau)$ in the ensemble $j_\varphi^\alpha(x,\tau)$ defined along the 
worldline, recovering the standard particle current through
\begin{equation}
\begin{array}{rcl}
j_\varphi^{\mu } \left( x,\tau \right)= 
\dint ds~\varphi \left( \tau -s\right)
j^{\mu }\left( x,s\right) &\longrightarrow& \dint ds~ 1 \cdot j^{\mu }\left( x,s\right) =
J^\mu (x) \strt{18}\\
j_\varphi^{5 } \left( x,\tau \right)= 
\dint ds~\varphi \left( \tau -s\right)
j^{5 }\left( x,s\right) &\longrightarrow& \dint ds~ j^{5 }\left( x,s\right)
 \strt{18}\\
\partial _{\mu } j_\varphi^{\mu }\left( x,\tau \right) +
\dfrac{1}{c_5}\partial _{\tau } j^5_\varphi \left( x,\tau \right)  = 0
&\longrightarrow& \partial _{\mu }
J^{\mu }\left( x \right) = 0
\end{array}
\end{equation}
of Maxwell theory.  {{Note that the current} $\dfrac{1}{c_5}j^5_\varphi \left( x,\tau
\right)$ {remains finite because }$j^5(x,\tau)$ {includes the factor}
$\dot X^5 = c_5$}. %%% footnote moved into paragraph
As expected, $f^{\mu\nu}$ decouples from $f^{5\mu}$ and satisfies Maxwell's
equations, while the photon mass $m_\gamma \sim \hbar / \xi \lambda c^2$
vanishes.
 
\section{Conclusions}

Stueckelberg--Horwitz--Piron electrodynamics can be approached as an abstract
gauge theory, exploring the consequences of allowing the gauge transformation
(\ref{wave}) of the quantum wave function to depend on the evolution parameter
in the dynamical framework.  However, in another sense, just~as Maxwell sought to
formalize the empirical results of Cavendish and Coulomb, SHP may be~seen as
accounting for classical Maxwell electrodynamics in light of the pair
creation/annihilation phenomena observed by Anderson.  For Stueckelberg, pair
processes provide empirical evidence that time must be~understood as two distinct
physical phenomena, chronology and coordinate, and so must be~formalized through
independent quantities $\tau$ and $(x^0,{\mathbf x})$ in a physically reasonable
theory.  Having~become accustomed, during the two hundred years that separate
Cavendish from general gauge theory, to~characterizing a single physical time by
the evolution of specialized machines (clocks) in~a~coordinate frame, 
it~is~unsurprising that introducing such a distinction raises conceptual difficulties. 

A new phenomenon in SHP is the absence of a static configuration---a particle
may only remain at the origin in its rest frame for all coordinate time $x^0$ if
its underlying microscopic event 
continually and uniformly evolves along its time axis, as $x = \left( c(\tau +
\tau_0)\strt{4},{\mathbf 0}\right) $.   The coordinate $x^0 = c\tau_0$ at $\tau
= 0$ is~not simply an artifact of initializing a system clock, because the field
induced by this event trajectory  
\begin{equation}
a^{0} (x,\tau) =  \frac{e}{{4\pi \vert \mathbf{x} \vert }}\varphi
\left(\tau + \tau_0 - \left( t - \frac{\vert \mathbf{x} \vert}{c} \right) \right)
\qquad a^{5}(x,\tau) = \frac{c_5}{c} a^{0} (x,\tau)
\label{yukawa-2}
\end{equation}
depends explicitly on the constant $\tau_0$.  The irreversible concatenation
performed by measuring apparatus recovers the familiar Coulomb potential
\begin{equation}
A^0(x) = \int \dfrac{d\tau}{\lambda} \; a^0(x,\tau) = \frac{e}{{4\pi \vert \mathbf{x} \vert }}
\label{cool}
\end{equation}
with no dependence on $\tau_0$ or even on the details of the weight function
$\varphi (\tau)$.  Similarly, $\tau_0$ plays no role in the quantized theory
where sharply defined mass-momentum states retain no information about 
the~initial conditions of coordinates.  Nevertheless, in SHP, 
the microscopic event dynamics are~determined by the Lorentz force (\ref{lorentz})
and so a test event at $x(\tau )=(c(\tau + \tau_0^\prime),\mathbf{x})$ will 
experience the Coulomb force 
\begin{equation}
M\mathbf{\ddot{x}}  
=-e^{2}~\frac{1-g_{55}\frac{c_{5}}{c}}{1+\left( \frac{c_{5}}{c}\right) ^{2}
}\nabla \left( \dfrac{e^{-\left\vert c(\tau_0 - \tau_0^\prime) + \left\vert \mathbf{x} \right\vert\right\vert
/\xi \lambda c}}{{ 4\pi }\left\vert \mathbf{x}\right\vert }\right)
\label{lor-coul-2}
\end{equation}
depending on the synchronization $\tau_0 - \tau_0^\prime$.  Regarding
(\ref{yukawa-2}) as a Yukawa potential, the limit $\lambda \rightarrow 0$ is
understood as extending the range of the interaction to a Coulomb form by taking the
mass of the photons that carry the interaction to zero.  Viewing $\lambda$ as a
correlation time for the microscopic current density $j^\alpha (x , \tau)$, this
limit can be understood as smoothing the interaction to a time-independent Coulomb form
by extending
the correlation between values of the potential at different times along the
entire worldline.

The structure of the source current for the pre-Maxwell field equations was seen
to be determined by the choice of kinetic term for the fields.  Standard field
theory texts note that this choice is not imposed by physical foundations, but recommend the
simplest form $f^{\alpha\beta}f_{\alpha\beta}$ because it is Lorentz and gauge
invariant, contains only first order derivatives, and in the case of Maxwell
theory, recovers the~known field equations.  For SHP, this choice is
equivalent to taking $\varphi (\tau) = \lambda \delta (\tau)$, which does recover the
concatenated Coulomb force through (\ref{cool}) but renders the
Lorentz force 
\begin{equation}
M\mathbf{\ddot{x}}  
=-\lambda e^{2}\left( 1-g_{55}\frac{c_{5}}{c}\right)
\nabla \left[ \dfrac{1}{{ 4\pi }\left\vert \mathbf{x}\right\vert }
\ \delta \left(\tau_0 - \tau_0^\prime + \left\vert \mathbf{x}\right\vert / c \right)\right]
\label{lor-coul-3}
\end{equation}
difficult, if not impossible, to reconcile with known
phenomenology.  The smooth (but non-local) current was found by adding the higher-derivative term
$ \left( \partial_\tau f^{\alpha \beta }\right) \left( \partial_\tau f_{\alpha
\beta }\right)$, which does not affect the~Lorentz and gauge invariance of the action.  
A term of this type has also been considered by Pavsic for brane interactions
\cite{pavsic}.

Limits on the values of the parameters in SHP can be found from standard phenomenology.
To~describe elastic particle-antiparticle scattering, (\ref{lor-coul-2}) undergoes
\begin{equation}
- e^{2}\left( 1-g_{55}\frac{c_{5}}{c}\right) \rightarrow
e^{2}\left( 1+g_{55}\frac{c_{5}}{c}\right)
\end{equation}
so that the experimental error in the asymmetry of scattering cross-sections 
places a limit on $c_5 / c \ll 1$ and allows us to take $\xi \sim
1/2$.  The width of the distribution $\varphi(\tau)$ is characterized by the
parameter $\lambda$, which also determines as $m_\gamma \sim \hbar / \xi \lambda
c^2 $ the mass spectrum of the photons that mediate the induced force.  Taking
$m_\gamma$ to be the experimental error on the mass of the photon ($10^{-18} eV/c^2$),
we may estimate $\lambda >
10^{-2}$ seconds.  Thus, only very low energy interactions will produce
phenomena that can be distinguished from standard Maxwell electrodynamics.  
Nevertheless, two distinct experimental signatures have been described.  It was
shown in \cite{super} that the Li\'{e}nard--Wiechert potential for a classical
linear particle trajectory experiencing virtual particle--antiparticle processes
will differ slightly from the~potential predicted by Maxwell theory.  In
addition, it was shown in \cite{qft} that the scattering cross-section for
scalar particles will differ slightly from the Klein--Nishina formula.

The smoothing of the single-event current by $\varphi(\tau)$ can be understood
as constructing a~statistical ensemble of events as the source for the field
equations.  An event on the trajectory $X^\mu(\tau)$ is~associated with an
ensemble whose members are of the form $\varphi(s) X^\mu(\tau-s)$, where the
weight $\varphi(s)$ is~the~probability that a process generating independent
random events at a constant average rate will produce an event occurring at
displacement $s$ from time $\tau$.  A single member of the ensemble is selected
by the causal properties of the Green's function when determining the potential
induced by the event trajectory.  In this way, classical statistical mechanics
is fundamental to the concept of a single-particle system and to the construction of a
well-posed relativistic Hamiltonian theory of~electromagnetism.

\section*{References}

%
%%%%%%%%%%%%%%%%%%%%%%%%%% REFERENCES %%%%%%%%%%%%%%%%%%%%%%%%%%

%
%
%
%

\end{document}